# Reliable and High Performance IGZO and In$_2$O$_3$ Transistors via Channel Capping

C. W. Cheng[1], J. Smith[1], K. Mashooq[2], P. Solomon[1], R. Watters[1], T. Philicelli[2], D. Piatek[1], C. Lavoie[1], M. Hopstaken[1], L. Gignac[1], B. Khan[1], M. BrightSky[1], G. Gionta[1], P. Hashemi[1], V. Narayanan[1], M. M. Frank[1]

[1]IBM Research, Yorktown Heights, NY, USA  [2]IBM Albany NanoTech, Albany, NY, USA, email: chengwei@us.ibm.com

## Abstract

A device and process strategy for achieving reliable indium gallium zinc oxide (IGZO) and indium oxide (InO) transistors compatible with a 400°C BEOL thermal budget and without performance degradation is demonstrated by fully exploiting intrinsic oxide material properties. An InO transistor with a novel amorphous In$_2$O$_3$–SiO$_2$ (InO-SiO) capping layer exhibits a positive threshold voltage (Vt), high extrinsic saturation mobility (33.1 cm$^2$V$^{-1}$s$^{-1}$), and only a 5mV Vt shift after positive-bias stress (PBS) at 3 MV/cm for 1000 s at room temperature, superior to conventional SiO$_2$ encapsulation.

## Introduction

IGZO and InO are promising materials for BEOL active devices. InO is particularly attractive due to its high mobility for high-speed electronics [1], but it can suffer from reliability and stability issues. Although an amorphous phase is preferred for lower device-to-device variability, thicker InO films tend to crystallize. Alloying or elemental doping is commonly used to improve reliability with constraint of InO thickness; however, mobility often degrades significantly before acceptable reliability is achieved (Fig. 1) [2-6]. This work proposes a strategy to achieve reliable oxide transistors without sacrificing performance by following intrinsic material properties of IGZO and InO.

## Experiments

Transistors were fabricated with well-designed shadow masks to ensure precise alignment of each active layer while avoiding extrinsic defectivity from lithographic patterning to show intrinsic material properties. Channel, source and drain (SD), capping, and encapsulation layers are deposited by sputtering. A 400°C anneal was performed to ensure devices pass criteria of BEOL thermal budget. Details of process, materials and device structure are shown in Fig. 2.

## Discussion

The equivalent circuit of a staggered oxide transistor and the relationship between depletion width and depletion gate voltage are illustrated in Fig. 3. Channel shutoff is governed by depletion, while current conduction under positive gate bias is dominated by the accumulation layer with some bulk layer contribution (mainly depends on carrier concentration in oxide layer). Adopting high-k gate dielectrics (higher gate capacitance) helps to reduce the operation voltage. Without heavily doped SD regions, access resistance of the oxide layer (r$_{acss}$) will dominate and impact the current significantly [7].

Thickness-dependent performance and hysteresis were observed in IGZO transistors while Vt is nearly thickness independent (Fig. 4). Thinner channels (smaller r$_{acss}$) yield higher ON current but worse hysteresis. Similar thickness trends were found in positive and negative bias stress (PBS/NBS) reliability for both IGZO and InO (Fig. 5). At identical thickness and stress (3 MV/cm), InO exhibits superior PBS stability compared to IGZO. Dielectrics passivation and encapsulation of oxide channel surface significantly improve PBS/NBS behavior (Fig. 6). Some phenomena were explained electrostatically by comparing the ratio of thickness and Debye length [8]. However, these results also indicate a strong contribution from material-dependent surface defect states on top of the oxide channel to device reliability. Adopting high-k gate dielectrics with higher gate capacitance (Cox) can further suppress Vt shift due to ΔVt = ΔQ / Cox. Improved reliability could be confirmed visually for the devices with same channel thickness but on high-k dielectrics (Fig.5(a)⚙ vs. Fig.6(b)⚙).

A fundamental tradeoff exists: thin channels are required for high performance, while thick channels are preferred for reliability. To overcome this dilemma, an innovative device structure and process flow for high performance and reliability were demonstrated to decouple different requirements on the thickness in channel and contact region. This innovative structure of IGZO transistor achieved by standard staggered device fabrication process flow but with additional oxide semiconductor material on top of the channel region before encapsulation is shown in Fig. 7. An innovative IGZO transistor with 10 nm channel contact achieves an extrinsic saturation mobility of 22 cm$^2$V$^{-1}$s$^{-1}$ with near-zero hysteresis and PBS shift of 15 mV even with 50 nm SiO$_2$ gate dielectrics (lower gate capacitance). To note, the thicker the channel region, the better the reliability results.

However, this approach cannot be directly applied to InO as (i) thick InO films tend to crystallize (Fig. 1b) and (ii) thin amorphous InO capped with other conventional amorphous semiconducting oxides (ex. IGZO, GaO, ZnO) can develop a high-carrier-concentration shunt path due to interdiffusion at InO/capping oxide interface (see Fig. 8a for IGZO-caped InO). This is caused by InGaO (Fig. 8b) or InZnO becoming highly conductive at specific compositions, resulting in the need for a high negative voltage to fully turn off the channel.

Preventing thick InO film (as a capping layer) from crystallization is another viable path. To realize this, $SiO_2$-doped InO was developed. Doping InO with $SiO_2$ was found to effectively increase its resistivity. With $SiO_2$ content above 25%, InO–SiO transitions into an amorphous insulating phase while maintaining compatibility with InO. InO–SiO can therefore function simultaneously as a virtual channel and encapsulation layer (Fig. 9).

Electrical characteristics and reliability of a 3.5 nm InO channel transistor with a InO-SiO capping layer on a 10 nm $HfO_2$ bottom gate dielectric are shown in Fig. 10. With appropriate annealing in oxygen, Vt could be tuned to be positive. The extracted saturation mobility is 33.1 $cm^2V^{-1}s^{-1}$, comparable to the mobility (34.5 $cm^2V^{-1}s^{-1}$) of an InO transistor without any encapsulation, with no degradation. Interestingly and in contrast, InO devices with conventional $SiO_2$ encapsulation show significantly reduced mobility (19 $cm^2V^{-1}s^{-1}$). Progress of PBS improvements could be seen with encapsulations and high-k dielectrics in Fig. 10(d). The best PBS result shows only a 5mV shift for devices with both InO-SiO capping and $HfO_2$ gate dielectrics. Further improvements could be achieved with thinner gate dielectrics, thicker InO-SiO capping, and tuning amorphous InO channel thickness.

Although the strategy and device data were demonstrated on long-channel devices, this strategy is extendable to short-channel transistors. InO-SiO is an insulator, and it can be deposited by sputtering or ALD and used similarly to $SiO_2$ for encapsulation on InO short channel device. For IGZO short channel device, the concern is the controllability of bottom gate to fully shut off the channel with extra IGZO capping layer between SD. Fig. 11 shows TCAD simulation of a 100 nm gate-length IGZO transistor with a 50 nm IGZO capping on 5 nm IGZO channel. The channel could be fully shut off like its long-channel counterpart.

## Conclusion

A material-driven strategy for achieving high-performance and reliable IGZO and InO transistors is proposed and demonstrated. InO–SiO is shown to be an effective encapsulation and virtual channel material, enabling InO devices with high mobility and excellent bias-stress stability without performance degradation. The approach is compatible with short-channel scaling and BEOL integration.

## Acknowledgments

The authors are grateful for support from Microelectronic Research Laboratory, Dr. N. Gong, Dr. D. Bishop, Dr. T.C. Chen, and Dr. H. Bu at IBM Research.

## Motivation and Challenge of Reliability Improvement

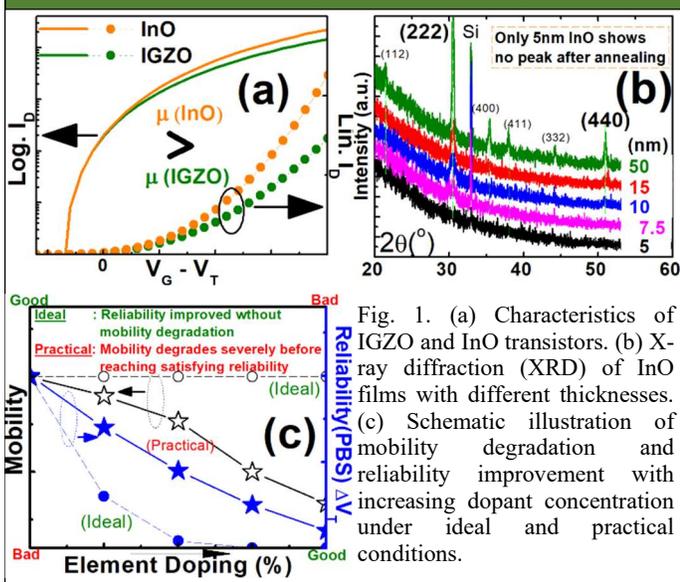

Fig. 1. (a) Characteristics of IGZO and InO transistors. (b) X-ray diffraction (XRD) of InO films with different thicknesses. (c) Schematic illustration of mobility degradation and reliability improvement with increasing dopant concentration under ideal and practical conditions.

## Experiments Details and Device Structure

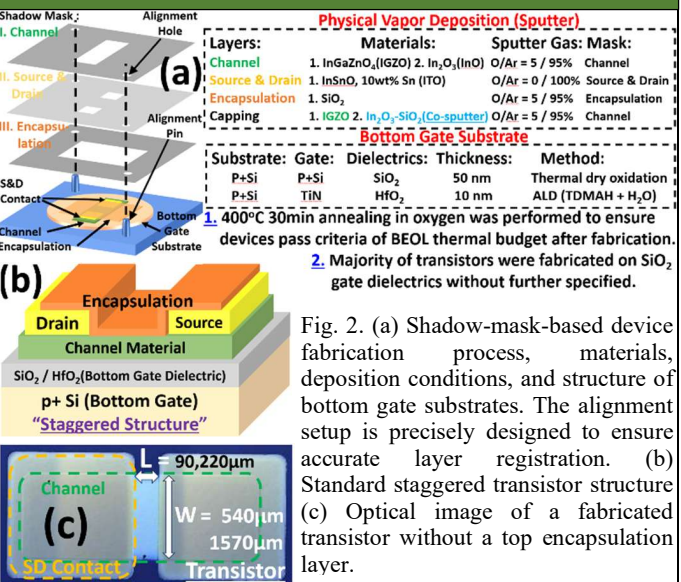

Fig. 2. (a) Shadow-mask-based device fabrication process, materials, deposition conditions, and structure of bottom gate substrates. The alignment setup is precisely designed to ensure accurate layer registration. (b) Standard staggered transistor structure (c) Optical image of a fabricated transistor without a top encapsulation layer.

## Fundamental Physics of Oxide Transistor

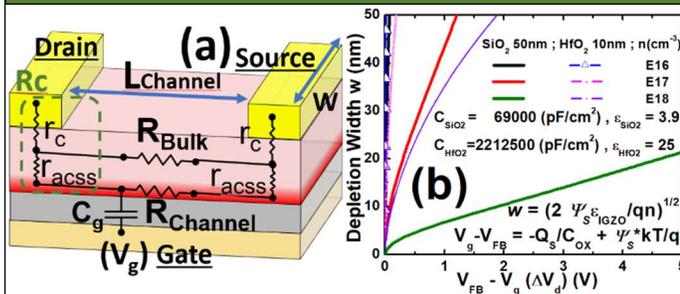

Fig. 3. (a) Equivalent circuit of staggered oxide transistor, where accumulation and bulk channels coexist. Contact resistance Rc consists of interfacial contact resistance $r_c$ and oxide film resistance $r_{acss}$, both dependent on carrier concentration and oxide channel thickness. Color contrast indicates carrier concentration distribution in channel. (b) Calculated depletion gate voltage ($\Delta V_d$) vs. depletion width (w) of carrier concentrations with 50nm $SiO_2$ and 10 nm $HfO_2$ as gate dielectrics. The channels of oxide FET are shut off by the depletion layer, where the depletion width is a function of carrier concentration, gate capacitance, and applied depletion gate voltage ($\Delta V_d$).

## Thickness Dependent Performance and Hysteresis

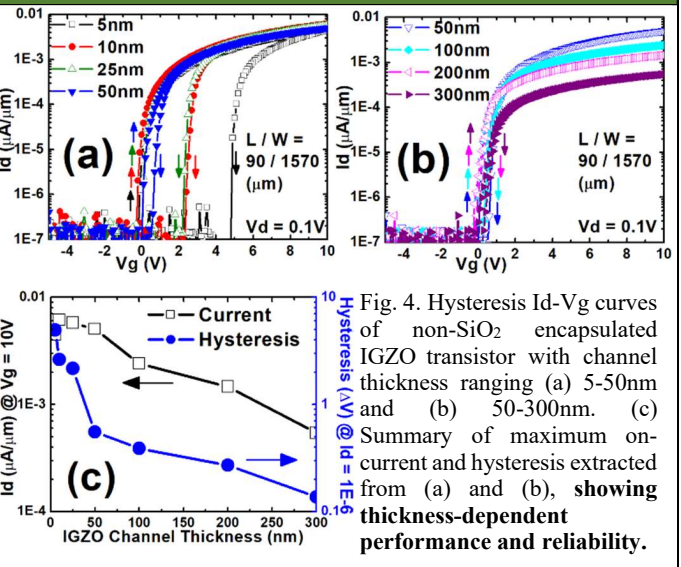

Fig. 4. Hysteresis Id-Vg curves of non-$SiO_2$ encapsulated IGZO transistor with channel thickness ranging (a) 5-50nm and (b) 50-300nm. (c) Summary of maximum on-current and hysteresis extracted from (a) and (b), **showing thickness-dependent performance and reliability.**

## Thickness Dependent IGZO and InO PBS and NBS

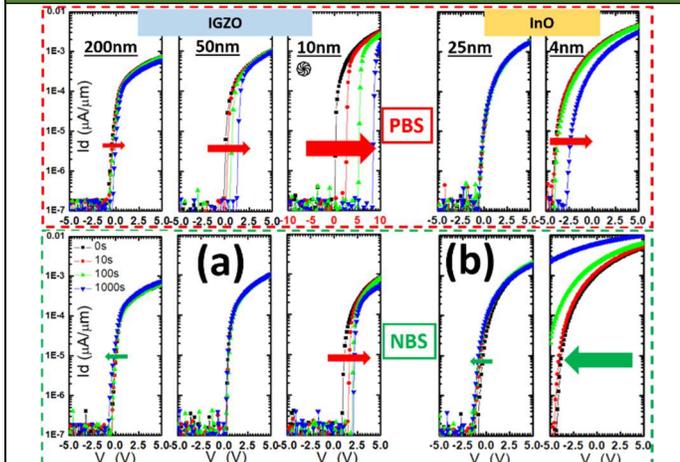

Fig. 5. PBS and NBS results of (a) 200, 50, and 10 nm IGZO transistors and (b) 25 and 4 nm InO transistor. All devices are without $SiO_2$ encapsulation (W/L =220/1570 μm). 3 MV/cm gate voltage stress were used for PBS/NBS (+/-15 V).

## Top Surface Passivation and High-K Gate Dielectrics

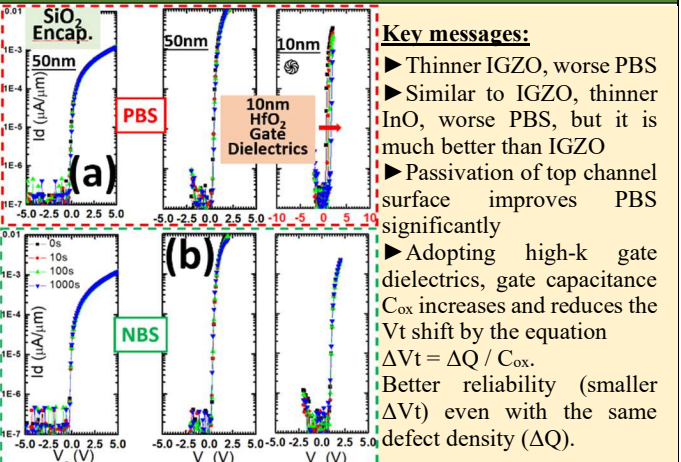

**Key messages:**
► Thinner IGZO, worse PBS
► Similar to IGZO, thinner InO, worse PBS, but it is much better than IGZO
► Passivation of top channel surface improves PBS significantly
► Adopting high-k gate dielectrics, gate capacitance $C_{ox}$ increases and reduces the Vt shift by the equation
$\Delta V_t = \Delta Q / C_{ox}$.
Better reliability (smaller $\Delta V_t$) even with the same defect density ($\Delta Q$).

Fig. 6. PBS and NBS results of (a) 50 nm IGZO transistors with $SiO_2$ encapsulation and (b) 50 nm and 10 nm IGZO transistor with 10 nm $HfO_2$ Gate Dielectrics (W/L =220/1570 μm). A gate electric field of 3 MV/cm is applied during stress.

## Innovative Structure of IGZO Transistor and Results

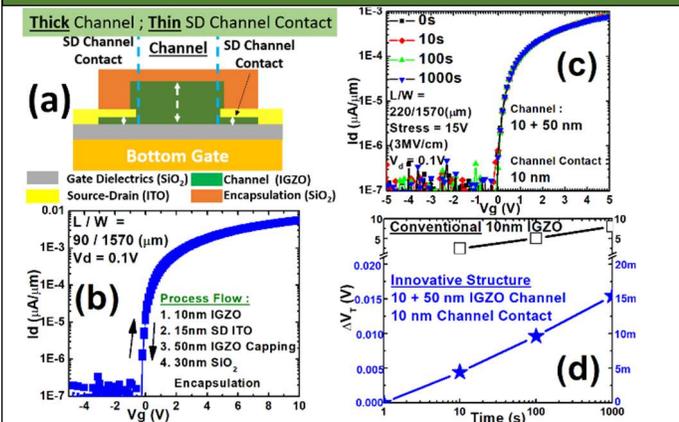

Fig. 7. (a) Schematic of the proposed innovative IGZO transistor structure. (b) Hysteresis of Id-Vg curves of 10 nm IGZO channel contact device (c) PBS characteristics of a 10-nm IGZO transistor with a 50-nm IGZO channel capping layer. (d) Comparison of $V_t$ shift under PBS between conventional and innovative 10 nm IGZO transistors.

## Challenge of InO Transistor with Extra Oxide Capping

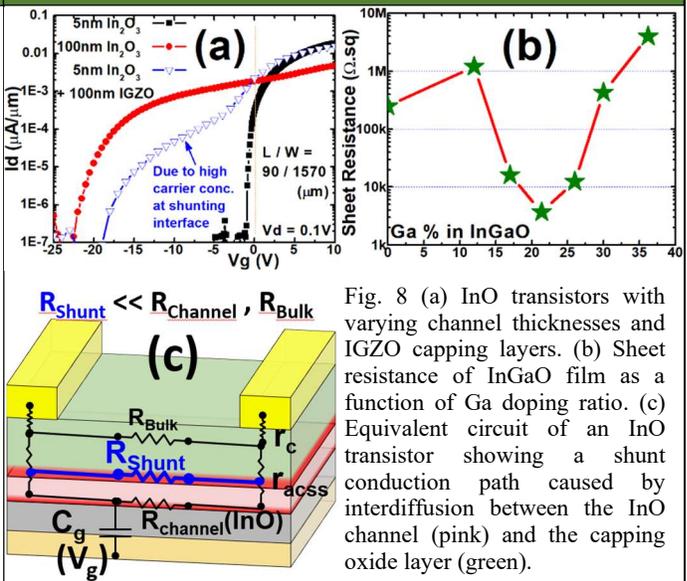

Fig. 8 (a) InO transistors with varying channel thicknesses and IGZO capping layers. (b) Sheet resistance of InGaO film as a function of Ga doping ratio. (c) Equivalent circuit of an InO transistor showing a shunt conduction path caused by interdiffusion between the InO channel (pink) and the capping oxide layer (green).

## Properties of InO-SiO and Proposed Device Structure

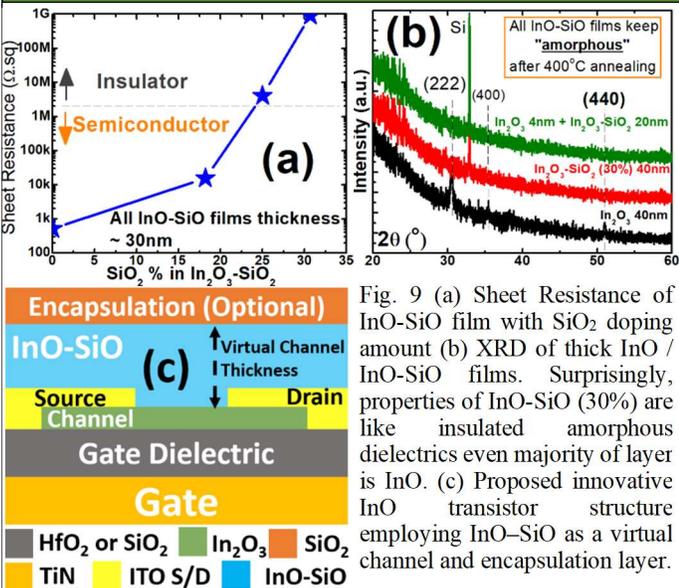

Fig. 9 (a) Sheet Resistance of InO-SiO film with $SiO_2$ doping amount (b) XRD of thick InO / InO-SiO films. Surprisingly, properties of InO-SiO (30%) are like insulated amorphous dielectrics even majority of layer is InO. (c) Proposed innovative InO transistor structure employing InO–SiO as a virtual channel and encapsulation layer.

## InO Transistor w. InO-SiO Capping and Results

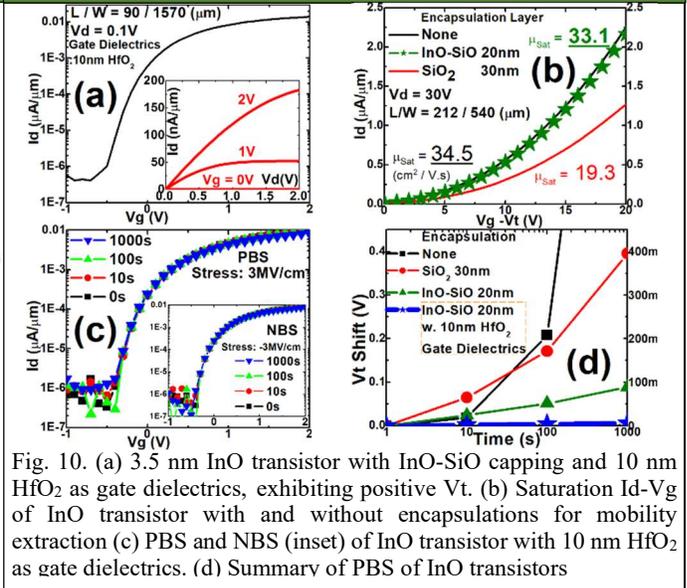

Fig. 10. (a) 3.5 nm InO transistor with InO-SiO capping and 10 nm $HfO_2$ as gate dielectrics, exhibiting positive Vt. (b) Saturation Id-Vg of InO transistor with and without encapsulations for mobility extraction (c) PBS and NBS (inset) of InO transistor with 10 nm $HfO_2$ as gate dielectrics. (d) Summary of PBS of InO transistors

## TCAD of Innovative IGZO Short Channel Device

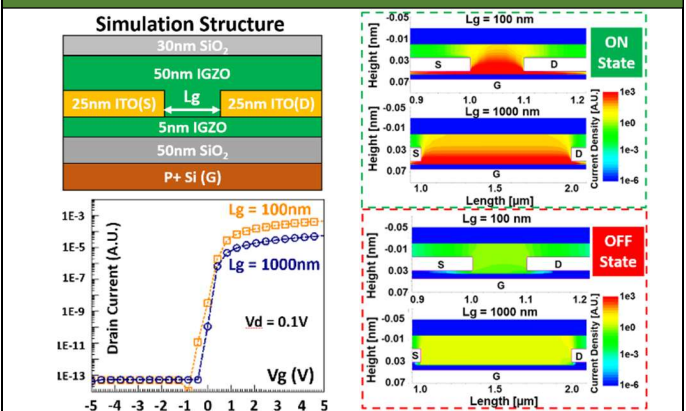

Fig. 11. TCAD simulation of IGZO transistors with innovative structure for short and long channel devices (Lg = 100 and 1000 nm), showing effective suppression of bulk channel conduction in OFF-State even with a relatively thick IGZO layer. Both accumulation and bulk current contribution can be observed in ON-State. IGZO bulk and surface defect density parameters in simulation [9]